\documentstyle[12pt]{article}

\global\arraycolsep=2pt
\makeatletter
\def\fmslash{\@ifnextchar[{\fmsl@sh}{\fmsl@sh[0mu]}}
\def\fmsl@sh[#1]#2{%
  \mathchoice
    {\@fmsl@sh\displaystyle{#1}{#2}}%
    {\@fmsl@sh\textstyle{#1}{#2}}%
    {\@fmsl@sh\scriptstyle{#1}{#2}}%
    {\@fmsl@sh\scriptscriptstyle{#1}{#2}}}
\def\@fmsl@sh#1#2#3{\m@th\ooalign{$\hfil#1\mkern#2/\hfil$\crcr$#1#3$}}
\makeatother

\begin{document}

\vspace{1.cm}
\begin{center}
\Large\bf Couplings of Pions with Excited Heavy Mesons From\\ 
Light-Cone QCD Sum Rules\\
 in the Leading Order of HQET 
\end{center}
\vspace{0.5cm}
\begin{center}
{Yuan-Ben Dai and Shi-Lin Zhu}\\ \vspace{3mm}
{\it Institute of Theoretical Physics,
 Academia Sinica, P.O.Box 2735, Beijing 100080, China }
\end{center}

\vspace{1.5cm}
\begin{abstract}
 The couplings of pions with lowest three doublets 
 $(0^-, 1^-)$, $(0^+,1^+)$ and $(1^+,2^+)$
of heavy mesons are studied with light-cone 
QCD sum rules in the leading order of heavy quark effective
theory. The ambiguity due to presence of two distinct $1^+$ states are solved.
\end{abstract}

{\large PACS number: 12.39.Hg, 14.40.Nd, 12.38.Lg}

\pagenumbering{arabic}

\section{ Introduction}
\label{sec1} 
Remarkbale progress has been made in understanding the physics of 
heavy mesons composed of a heavy quark and a light quark with the 
discovery of the heavy quark symmetry.
To obtain detailed predictions one has to employ some specific 
nonperturbative methods, among which QCD sum rules \cite{svz} is very useful.
The heavy quark effective theory (HQET) \cite{grinstein} provides a 
systematic expansion of QCD 
in terms of $1/ m_Q$, where $m_Q$ is the heavy quark mass.
The spectrum of the ground state
heavy meson has been studied with the QCD sum rules in HQET in \cite{col91}.
In \cite{huang} the mass of the lowest excited heavy meson 
doublets $(2^+,1^+)$ and $(1^+,0^+)$ were studied with QCD sum rules in the
heavy quark effective theory (HQET) up to the order of ${\cal O}(1/m_Q)$. 

QCD sum rules has been used to analyse the couplings of the 
heavy hadrons with pions [5-15]. The widths for pionic decays of the
lowest two excited doublets $(2^+,1^+)$ and $(1^+,0^+)$ is calculated 
with conventional QCD sum rules in the leading order of HQET in \cite{ybd97}.
Light-cone QCD sum rules (LCQSR) with the full QCD Lagrangian is 
first employed to derive the strong 
coupling constants: $g_{\pi D^{\ast} D}$, $g_{\pi B^{\ast} B}$ \cite{bely95}. 
Recently the strong coupling constants: $g_{\pi B^{\ast} B^{\ast} }$, 
$g_{\pi B^{\prime}_1 B^{\prime}_0}$ and $g_{\pi B^{\prime}_1 B^{\ast} }$ were calculated using 
LCQSR with finite heavy quark mass \cite{bari97}.
The couplings of heavy baryons with 
soft pions have been estimated with QCD sum rules in an external axial field \cite{grozin97}.
In this work we employ the LCQSR in HQET to calculate the on-shell couplings of the 
pions with heavy meson doublets $(0^-, 1^-)$, $(0^+,1^+)$ and $(1^+,2^+)$
to the leading order of $1/ m_Q$. 

The LCQSR is quite different from the conventional QCD sum rules, which is 
based on the short-distance operator product expansion (OPE). 
The LCQSR is based on the OPE on the light cone, 
which is the expansion over the twists of the operators. The main contribution
comes from the lowest twist operator. Matrix elements of nonlocal operators 
sandwiched between a hadronic state and the vacuum defines the hadron wave
functions. When the LCQSR is used to calculate the coupling constant, the 
double Borel transformation is always invoked so that the excited states and 
the continuum contribution can be treated quite nicely. Moreover, the final 
sum rule depends only on the value of the wave function at a specific point like
$\varphi_{\pi}(u_0 ={1\over 2})$, which is much better known than the whole wave 
function \cite{bely95}.

One difficult problem encountered in studying the decay widths of excited heavy mesons
with QCD sum rules is the following. Except for the lowest states $0^-$, 
$1^- $, the spectra contains a pair of states for any spin-parity $j^P$ with
close values in their masses but quite different in magnitudes of their
decay widths. In our case, one of the two $1^+$ states is a narrow resonance
decaying mainly by emitting a $D$ wave pion, while the other one is a very
wide resonance decaying by emitting a $S$ wave pion. An interpolating
current used for the narrow $1^+$ state with a small coupling to the other $%
1^+$ state may cause sizable error in the result of calculation. It is only
in the $m_Q\to\infty$ limit, there is a conserved quantum number $j_{\ell}$,
the angular momentum of the light component, which can be used to
differentiate the two states. Therefore, HQET has important and unique 
advantage for this purpose. These are the motivation for our approach of 
using LCQSR in HQET.

The proper interpolating current $J_{j,P,j_{\ell}}^{\alpha_1\cdots\alpha_j}$
for the states with the quantum number $j$, $P$, $j_{\ell}$ in HQET was
given in \cite{huang}. They were proved to satisfy the following conditions 
\begin{eqnarray}
\label{decay}
\langle 0|J_{j,P,j_{\ell}}^{\alpha_1\cdots\alpha_j}(0)|j',P',j_{\ell}^{'}\rangle&=&
f_{Pj_l}\delta_{jj'}
\delta_{PP'}\delta_{j_{\ell}j_{\ell}^{'}}\eta^{\alpha_1\cdots\alpha_j}\;,\\
\label{corr}
i\:\langle 0|T\left (J_{j,P,j_{\ell}}^{\alpha_1\cdots\alpha_j}(x)J_{j',P',j_{\ell}'}^{\dag
\beta_1\cdots\beta_{j'}}(0)\right )|0\rangle&=&\delta_{jj'}\delta_{PP'}\delta_{j_{\ell}j_{\ell}'}
(-1)^j\:{\cal S}\:g_t^{\alpha_1\beta_1}\cdots g_t^{\alpha_j\beta_j}\nonumber\\[2mm]&&\times\:
\int \,dt\delta(x-vt)\:\Pi_{P,j_{\ell}}(x)
\end{eqnarray}
in the $m_Q\to\infty$ limit, where $\eta^{\alpha_1\cdots\alpha_j}$ is the
polarization tensor for the spin $j$ state, $v$ is the velocity of the heavy
quark, $g_t^{\alpha\beta}=g^{\alpha\beta}-v^{\alpha}v^{\beta}$ is the
transverse metric tensor, ${\cal S}$ denotes symmetrizing the indices and
subtracting the trace terms separately in the sets $(\alpha_1\cdots\alpha_j)$
and $(\beta_1\cdots\beta_{j})$, $f_{P,j_{\ell}}$ and $\Pi_{P,j_{\ell}}$ are
a constant and a function of $x$ respectively which depend only on $P$ and $%
j_{\ell}$. Because of equations (\ref{decay}) and (\ref{corr}), the sum rule
in HQET for decay widths derived from a correlator containing such currents
receive no contribution from the unwanted states with the same spin-parity
as the states under consideration in the $m_Q\to\infty$. Starting from the
calculations in the leading order, the decay amplitudes for finite $m_Q$ can
be calculated unambiguously order by order in the $1/m_Q$ expansion in HQET.


\section{ Sum rules for decay amplitudes}

In the present work we shall confine ourselves to the lowest lying 
states in the leading order of $1/m_Q$ expansion. Denote the doublet $%
(1^+,2^+)$ with $j_{\ell}=3/2$ by $(B_1,B_2^*)$ and the doublet $(0^+,1^+)$
with $j_{\ell}=1/2$ by $(B^{\prime}_0,B^{\prime}_1)$. 
There are many combinations for the coupling constant
$g_{\pi B_x B_y}$, where $B_x$, $B_y$ belongs to the three doublets 
$(0^-, 1^-)$, $(0^+,1^+)$ and $(1^+,2^+)$. Due to the heavy quark symmetry 
and chiral symmetry there exist only six independent coupling constants, 
i.e., three independent coupling constants
when $B_x$, $B_y$ belongs to different doublets and another three
when $B_x$, $B_y$ belongs to the same doublet. For example, from covariance 
and conservation of the angular momentum of the light component in the $%
m_Q\to\infty$ limit, the amplitudes for the decay of $B_1$, $B_2^*$ to the
ground states $B$, $B^*$ in the doublet $(0^-,1^-)$ with $j_{\ell}=1/2$ have
the following forms 
\begin{eqnarray}
\label{coup1}
 M(B_1\to B^*\pi)&=&I\;\epsilon^*_{\mu}\eta_{\nu}(q_t^{\mu}q_t^{\nu}-\frac{1}{3}g_t^{\mu\nu}
 q_t^2)g(B_1,B^*)\;,\\
\label{coup2}
 M(B_2^*\to B\pi)&=&I\;\eta_{\mu\nu}q_t^{\mu}q_t^{\nu}g(B_2^*,B)\;,\\
 \label{coup3}
 M(B_2^*\to B^*\pi)&=&I\;i\varepsilon_{\alpha\beta\sigma\rho}\;\epsilon^{*\alpha}v^{\beta}
 \eta^{\sigma\mu}q_t^{\rho}q_{t\mu}g(B_2^*,B^*)\;,
\end{eqnarray} 
where $\eta_{\mu\nu}$, $\eta_{\mu}$ and $\epsilon_{\mu}$ are polarization
tensors for states $2^+$, $1^+$ and $1^-$ respectively. $q_{t\mu}=q_{\mu}-v%
\cdot qv_{\mu}$. $I=\sqrt{2}$, $1$ for charged and neutral pion
respectively. As shown in \cite{falk} the constants in 
(\ref{coup1})-(\ref{coup3}) satisfy 
\begin{eqnarray}
\label{ggg}
g(B_2^*,B)=g(B_2^*,B^*)=\sqrt{\frac{2}{3}}\;g(B_1,B^*)\;.
\end{eqnarray} 

We calculate the following six independent coupling constants:
$g_1=g_{\pi B_1 B^\ast}$, $g_2=g_{\pi B_0^{\prime} B}$, 
$g_3=g_{\pi B_1^{\prime} B_0^{\prime}}$, $g_4=g_{\pi B_2^{\ast} B_1}$,
$g_5=g_{\pi B_1^{\prime} B_1}$ and $g_6=g_{\pi B^{\ast} B}$.
Other coupling constants are related to them. 
There are related results for $g_2$, $g_3$ and 
$g_6$ in [6-15]. In \cite{ybd97} $g_1$ and $g_2$ is calculated with 
QCD sum rules in the short distance expansion. 
For the sake of comparison and completeness, we also present the sum rules for 
$g_2$, $g_3$ and $g_6$. 

For deriving the sum rules for the coupling constants we consider the correlators 
\begin{eqnarray}
\label{7a}
 \int d^4x\;e^{-ik\cdot x}\langle\pi(q)|T\left(J^{\beta}_{1,-,\frac{1}{2}}(0)
 J^{\dagger\alpha}_{1,+,\frac{3}{2}}(x)\right)|0\rangle = 
 \left(q_t^{\alpha}q_t^{\beta}-\frac{1}{3}g_t^{\alpha\beta}q^2_t\right)I\;G_{B_1B^*}
 (\omega,\omega')\;,\\[2mm]\label{7b}
 \int d^4x\;e^{-ik\cdot x}\langle\pi(q)|T\left(J_{0,+,\frac{1}{2}}(0)
 J^{\dagger}_{0,-,\frac{1}{2}}(x)\right)|0\rangle=
I\; G_{B_0^{\prime} B} (\omega,\omega')\;,\hspace{1.4cm}\\[2mm]\label{7c}
\int d^4x\;e^{-ik\cdot x}\langle\pi(q)|T\left(J_{0,+,\frac{1}{2}}(0)
 J^{\dagger\alpha}_{1,+,\frac{1}{2}}(x)\right)|0\rangle=
 q_t^{\alpha} I\;G_{B_1^{\prime}B_0^{\prime}} (\omega,\omega')\;,
 \\[2mm]\label{7d}\nonumber
\int d^4x\;e^{-ik\cdot x}\langle\pi(q)|T\left(J^{\alpha}_{1,+,\frac{3}{2}}(0)
 J^{\dagger\sigma\beta}_{2,+,\frac{3}{2}}(x)\right)|0\rangle= \\  
 \left(q_t^{\alpha}q_t^{\sigma}q_t^{\beta}-
 \frac{1}{6}q^2_t(g_t^{\alpha\beta}q_t^{\sigma} +
 g_t^{\alpha\sigma}q_t^{\beta} +
 {4\over 3}g_t^{\sigma\beta}q_t^{\alpha} )
 \right)
I\; G_{B_2^*B_1} (\omega,\omega')\;,\\[2mm]\label{7e}
\int d^4x\;e^{-ik\cdot x}\langle\pi(q)|T\left(J^{\alpha}_{1,+,\frac{3}{2}}(0)
 J^{\dagger\beta}_{1,+,\frac{1}{2}}(x)\right)|0\rangle=
 i\epsilon^{\alpha\beta\sigma\delta} q^t_{\sigma}v_{\delta}
I\; G_{B_1^{\prime}B_1} (\omega,\omega')\;,\\[2mm]\label{7f}
 \int d^4x\;e^{-ik\cdot x}\langle\pi(q)|T\left(J^{\alpha}_{1,-,\frac{1}{2}}(0)
 J^{\dagger}_{0,-,\frac{1}{2}}(x)\right)|0\rangle
 = q_t^{\alpha} I\;G_{B^*B}(\omega,\omega')\;,
\end{eqnarray}
where $k^{\prime}=k+q$, $\omega=2v\cdot k$, $\omega^{\prime}=2v\cdot
k^{\prime}$ and $q^2=0$. 
The forms of the right hand side of (\ref{7a})-(\ref{7f}) 
are determined by 
the conservation of angular momentum of the light component and the fact 
that $\alpha$, $\beta$, $\sigma$ and $\delta$ are transverse
indices, $x-y=vt$ on the heavy quark propagator.
The interpolationg currents are given in \cite{huang} as 
\begin{eqnarray}
\label{curr1}
&&J^{\dag\alpha}_{1,+,{3\over 2}}=\sqrt{\frac{3}{4}}\:\bar h_v\gamma^5(-i)\left(
{\cal D}_t^{\alpha}-\frac{1}{3}\gamma_t^{\alpha}\fmslash{\cal D}_t\right)q\;,\\
\label{curr2}
&&J^{\dag\alpha_1,\alpha_2}_{2,+,{3\over 2}}=\sqrt{\frac{1}{2}}\:\bar h_v
\frac{(-i)}{2}\left(\gamma_t^{\alpha_1}{\cal D}_t^{\alpha_2}+
\gamma_t^{\alpha_2}{\cal D}_t^{\alpha_1}-{2\over 3}g_t^{\alpha_1\alpha_2}
\fmslash{\cal D}_t\right)q\;,\\
\label{curr3}
&&J^{\dag\alpha}_{1,-,{1\over 2}}=\sqrt{\frac{1}{2}}\:\bar h_v\gamma_t^{\alpha}
q\;,\hspace{1.5cm} J^{\dag\alpha}_{0,-,{1\over 2}}=\sqrt{\frac{1}{2}}\:\bar h_v\gamma_5q\;,
\end{eqnarray}
\begin{eqnarray}
\label{current1}
J^{\dag}_{0,+,{1\over 2}}=\frac{1}{\sqrt{2}}\:\bar h_vq\;,\hspace{1.5cm}
J^{\dag\alpha}_{1,+,{1\over 2}}=\frac{1}{\sqrt{2}}\:\bar h_v\gamma^5\gamma^{\alpha}_tq\;.
\end{eqnarray} 
where $h_v$ is the heavy quark field in HQET and $\gamma_{t\mu}=\gamma_%
\mu-v_\mu\fmslash v$.

Let us first consider the function $G_{B_1B^*}(\omega,\omega^{\prime})$ in (%
\ref{7a}). As a function of two variables, it has the following pole terms
from double dispersion relation 
\begin{eqnarray}
\label{pole}
{f_{-,{1\over 2}}f_{+,{3\over 2}}g(B_1B^*)\over (2\bar\Lambda_{-,{1\over 2}}
-\omega')(2\bar\Lambda_{+,{3\over 2}}-\omega)}+{c\over 2\bar\Lambda_{-,{1\over 2}}
-\omega'}+{c'\over 2\bar\Lambda_{+,{3\over 2}}-\omega}\;,
\end{eqnarray}
where $f_{P,j_\ell}$ are constants defined in (\ref{decay}), $%
\bar\Lambda_{P,j_\ell}=m_{P,j_\ell}-m_Q$. As explained in Section \ref{sec1}%
, only one state with $j^P=1^+$ contributes to (\ref{pole}) as the result of
equation (\ref{decay}). This would not be true if the last term in (\ref
{curr1}) is absent.

For deriving QCD sum rules we calculate the correlator (\ref{7a}) by the
operator expansion on the light-cone in HQET to the leading 
order of ${\cal O}(1/ m_Q)$. 
The expression for $G_{B_1 B^{\ast}}(\omega, \omega')$ reads
\begin{equation}\label{lam5}			
{\sqrt{6}\over 8} \int_0^{\infty} dt \int dx e^{-ikx} 
\delta (-x-vt){\bf Tr} \{ 
 \gamma^t_\beta (1+{\hat v})\gamma_5 (D^t_\alpha 
 -{1\over 3}\gamma^t_\alpha {\hat D}^t )
\langle \pi (q)|u(0) {\bar d}(x) |0\rangle  
 \} \; ,
\end{equation}
The pion wave function is 
defined as the matrix elements of nonlocal operators between the vacuum and 
pion state. Up to twist four they are
\cite{bely95}:
\begin{eqnarray}
<\pi(q)| {\bar d} (x) \gamma_{\mu} \gamma_5 u(0) |0>&=&-i f_{\pi} q_{\mu} 
\int_0^1 du \; e^{iuqx} (\varphi_{\pi}(u) +x^2 g_1(u) + {\cal O}(x^4) ) 
\nonumber \\
&+& f_\pi \big( x_\mu - {x^2 q_\mu \over q x} \big) 
\int_0^1 du \; e^{iuqx}  g_2(u) \hskip 3 pt  , \label{ax} \\
<\pi(q)| {\bar d} (x) i \gamma_5 u(0) |0> &=& {f_{\pi} m_{\pi}^2 \over m_u+m_d} 
\int_0^1 du \; e^{iuqx} \varphi_P(u)  \hskip 3 pt ,
 \label{pscal}  \\
<\pi(q)| {\bar d} (x) \sigma_{\mu \nu} \gamma_5 u(0) |0> &=&i(q_\mu x_\nu-q_\nu 
x_\mu)  {f_{\pi} m_{\pi}^2 \over 6 (m_u+m_d)} 
\int_0^1 du \; e^{iuqx} \varphi_\sigma(u)  \hskip 3 pt .
 \label{psigma}
\end{eqnarray}
The wave function $\varphi_{\pi}$ is associated with the leading twist two 
operator, $g_1$ and $g_2$ correspond to twist four operators, and $\varphi_P$ and 
$\varphi_\sigma$ to twist three ones. Due to the choice of the
gauge  $x^\mu A_\mu(x) =0$, the path-ordered gauge factor
$P \exp\big(i g_s \int_0^1 du x^\mu A_\mu(u x) \big)$ has been omitted.

Expressing (\ref{lam5}) with the pion light-cone wave functions, we arrive at
\begin{equation}\label{q1}
G_{B_1 B^{\ast}}(\omega, \omega')= 
-i{\sqrt{6}\over 8}F_\pi\int_0^{\infty} dt \int_0^1 du e^{i (1-u) {\omega t \over 2}}
e^{i u {\omega' t \over 2}} u \{ \varphi_{\pi} (u) +
t^2g_1(u)+{it\over q\cdot v}g_2(u)+{it\over 6}\mu_{\pi}\varphi_{\sigma}(u) \}
+\cdots\; ,
\end{equation}
where $\mu_{\pi}\equiv {m_\pi^2 \over m_u+m_d}=1.76$GeV,  
$F_{\pi}={f_\pi\over \sqrt{2}}=92$MeV for neutral pions. 
For large euclidean values of $\omega$ and $\omega'$ 
this integral is dominated by the region of small $t$, therefore it can be 
approximated by the first a few terms.

Similarly, we have:
\begin{equation}\label{q2}
G_{B_0^{\prime} B} (\omega,\omega')
={i\over 4}F_\pi\int_0^{\infty} dt \int_0^1 du e^{i (1-u) {\omega t \over 2}}
e^{i u {\omega' t \over 2}}  \{\mu_\pi \varphi_P (u) -(q\cdot v)[
\varphi_\pi (u) + t^2g_1(u)]
\}\; .
\end{equation}
\begin{equation}\label{q3}
G_{B_1^{\prime}B_0^{\prime}} (\omega,\omega')=
{i\over 4}F_\pi\int_0^{\infty} dt \int_0^1 du e^{i (1-u) {\omega t \over 2}}
e^{i u {\omega' t \over 2}} \{ \varphi_{\pi} (u) +
t^2g_1(u)+{it\over q\cdot v}g_2(u)-{it\over 6}\mu_{\pi}\varphi_{\sigma}(u) \}
\; ,
\end{equation}
\begin{equation}\label{q4}
G_{B_2^*B_1} (\omega,\omega')=
-i{\sqrt{6}\over 8}F_\pi\int_0^{\infty} dt \int_0^1 du e^{i (1-u) {\omega t \over 2}}
e^{i u {\omega' t \over 2}} u^2 \{ \varphi_{\pi} (u) +
t^2g_1(u)+{it\over q\cdot v}g_2(u)+{it\over 6}\mu_{\pi}\varphi_{\sigma}(u) \}
\;,
\end{equation}
\begin{equation}\label{q5}
G_{B_1^{\prime}B_1} (\omega,\omega')=
i{\sqrt{6}\over 24}F_\pi\int_0^{\infty} dt \int_0^1 du e^{i (1-u) {\omega t \over 2}}
e^{i u {\omega' t \over 2}}  u \{\mu_\pi \varphi_P (u) +(q\cdot v)[
\varphi_\pi (u) + t^2g_1(u)]
\}\; ,
\end{equation}
\begin{equation}\label{q6}
G_{B^*B}(\omega,\omega')
={i\over 4}F_\pi\int_0^{\infty} dt \int_0^1 du e^{i (1-u) {\omega t \over 2}}
e^{i u {\omega' t \over 2}} \{ \varphi_{\pi} (u) +
t^2g_1(u)+{it\over q\cdot v}g_2(u)+{it\over 6}\mu_{\pi}\varphi_{\sigma}(u) \}
\;.
\end{equation}

After performing Wick rotation and double Borel transformation 
with the variables $\omega$ and $\omega'$
the single-pole terms in (\ref{pole}) are eliminated. 
Subtracting the continuum contribution which is modeled by the integral in region 
$\omega ,\omega' \ge \omega_c$, 
we arrive at:
\begin{equation}\label{final-a}
 g_1 f_{-,{1\over 2} } f_{+, {3\over 2} } = -{\sqrt{6}\over 4}F_{\pi}
 e^{ { \Lambda_{-,{1\over 2} } +\Lambda_{+,{3\over 2} } \over T }}
 \{ u_0\varphi_{\pi} (u_0)Tf_0({\omega_c\over T}) -
{4\over T}u_0g_1(u_0)+{4\over T}G_1(u_0)
+{1\over 3}\mu_{\pi}u_0\varphi_{\sigma}(u_0) \}\;,
\end{equation}
where $u_0={T_1\over T_1+T_2}$, $T={T_1T_2\over T_1+T_2}$,
$T_1$, $T_2$ are the Borel parameters,
$G_1 (u_0)\equiv \int_0^{u_0} ug_2(u)du$ and
$f_n(x)=1-e^{-x}\sum\limits_{k=0}^{n}{x^k\over k!}$. The presence of the factor 
 $f_n$ is the result of subtracting the integral 
$\int_{\omega_c}^\infty s^n e^{-{s\over T}} ds$ as a contribution of the continum.
In obtaining (\ref{final-a}) we have used the Borel transformation formula:
${\hat {\cal B}}^T_{\omega} e^{\alpha \omega}=\delta (\alpha -{1\over T})$.
The integral in the function $G_1 (u_0)$ in (\ref{final-a}) arises from 
the factor $1/(q\cdot v)$ in (\ref{q1}). Here we have used integration by parts 
to absorb the factor $1/(q\cdot v)$. In this way we 
arrive at the simple form after double Borel transformation. In the following we 
shall use the same technique to deal with the factor $(q\cdot v)^{-1}$ in other sum
rules.

Similarly we have:
\begin{equation}\label{final-b}
 g_2 f_{-,{1\over 2} } f_{+, {1\over 2} } =  {1\over 4}F_{\pi}
 e^{ { \Lambda_{-,{1\over 2} } +\Lambda_{+,{1\over 2} } \over T }}
 \{ -\varphi_\pi^{\prime} (u_0) T^2f_1({\omega_c\over T}) +
2\mu_\pi \varphi_P (u_0)T f_0({\omega_c\over T}) +4 g'_1(u_0)
\}\;,
\end{equation}
where $\varphi_\pi^{\prime} (u_0)$, $g'_1(u_0)$ are the first derivatives of
$\varphi_\pi (u)$, $g_1(u)$ at $u=u_0$. The appearance of derivatives is due to the 
factor $(q\cdot v)$.

\begin{equation}\label{final-c}
 g_3 f^2_{+,{1\over 2} }  = {1\over 2}F_{\pi}
 e^{ {2 \Lambda_{+,{1\over 2} }  \over T }}
 \{ \varphi_{\pi} (u_0)Tf_0({\omega_c\over T}) -
{4\over T}g_1(u_0)+{4\over T}G_3(u_0)
-{1\over 3}\mu_{\pi}\varphi_{\sigma}(u_0) \}\;,
\end{equation}
where $G_3 (u_0)\equiv \int_0^{u_0} g_2(u)du$.

\begin{equation}\label{final-d}
 g_4 f^2_{+,{3\over 2} }  = {\sqrt{6}\over 4}F_{\pi}
 e^{ 2\Lambda_{+,{3\over 2} } \over T }
 \{ u_0^2\varphi_{\pi} (u_0)Tf_0({\omega_c\over T}) -
{4\over T}u_0^2g_1(u_0)+{4\over T}G_4(u_0)
+{1\over 3}\mu_{\pi}u_0^2\varphi_{\sigma}(u_0) \}\;,
\end{equation}
where $G_4 (u_0)\equiv \int_0^{u_0} u^2g_2(u)du$.

\begin{equation}\label{final-e}
 g_5 f_{+,{1\over 2} } f_{+, {3\over 2} } = {\sqrt{6}\over 12}F_{\pi}
 e^{ { \Lambda_{+,{1\over 2} } +\Lambda_{+,{3\over 2} } \over T }}
 \{ {d\over du}\left( u\varphi_\pi (u) \right)|_{u_0}
  T^2f_1({\omega_c\over T}) +
\mu_\pi u_0\varphi_P (u_0)T f_0({\omega_c\over T}) -
4 {d\over du}\left( ug_1(u) \right)|_{u_0} 
\} \; ,
\end{equation}

\begin{equation}\label{final-f}
 g_6 f^2_{-,{1\over 2} }  = {1\over 2}F_{\pi}
 e^{ 2 \Lambda_{-,{1\over 2} }  \over T }
 \{ \varphi_{\pi} (u_0)Tf_0({\omega_c\over T}) -
{4\over T}g_1(u_0)+{4\over T}G_3(u_0)
+{1\over 3}\mu_{\pi}\varphi_{\sigma}(u_0) \} \;.
\end{equation}

\section{Determination of the parameters}

\label{sec3} 
In order to obtain the coupling constants 
from (\ref{final-a})-(\ref{final-f}) we need to use
the mass parameters $\bar\Lambda$'s and the coupling constants $f$'s of the
corresponding interpolating currents as input. $\bar\Lambda_{-,1/2}$ and $%
f_{-,1/2}$ can be obtained from the results in \cite{neubert} as $%
\bar\Lambda_{-,1/2}=0.5$ GeV and 
$f_{-,1/2}\simeq 0.25$ GeV$^{3/2}$ at the
order $\alpha_s=0$. Notice that the coupling constant $f_{-,1/2}$ defined in
the present work is a factor $1/\sqrt 2$ smaller than that defined in \cite
{neubert}. $\bar\Lambda_{+,3/2}$ and $\bar\Lambda_{+,1/2}$ 
are given in \cite{huang}. $f_{+,3/2}$ and $f_{+,1/2}$
can be determined from the formulas
(34), (27) and (28) of reference \cite{huang} derived from sum rules for two
point correlators. The results are 
\begin{eqnarray}
\label{fvalue}
&&\bar\Lambda_{+,3/2}=0.82 ~~\mbox{GeV}\hspace{1.2cm}f_{+,3/2}=0.19\pm 0.03 ~~\mbox{GeV}^{5/2}\;,\nonumber\\
&&\bar\Lambda_{+,1/2}=1.15 ~~\mbox{GeV}\hspace{1.1cm}f_{+,1/2}=0.40\pm 0.06 ~~\mbox{GeV}^{3/2}\;.
\end{eqnarray}

We use the wave functions adopted in \cite{bely95} to compute 
the coupling constants.
Moreover, we choose to work at the symmetric point $T_1=T_2=2T$, i.e., 
$u_0 ={1\over 2}$ as traditionally done in literature \cite{bely95}. 
We adopt the scale $\mu =1.4$GeV, at which the values of the various 
functions appearing 
in (\ref{final-a})-(\ref{final-f}), at $u_0={1\over2}$, are: 
$\varphi_\pi(u_0)=1.22\pm 0.3$,
$\varphi_P(u_0)=1.142$, 
$\varphi_\sigma(u_0)=1.463$, 
$g_1(u_0)=3.4 \times 10^{-2} \; $GeV$^2$, 
$G_1 (u_0)=-4.5 \times 10^{-3} \; $GeV$^2$, 
$G_3 (u_0)=-2.0 \times 10^{-2} \; $GeV$^2$, 
$G_4 (u_0)=-1.3 \times 10^{-3} \; $GeV$^2$, 
$g_2 ({1\over 2}) =0$, 
$\varphi_\pi^{\prime} (u_0)=0$
and $g'_1(u_0)=0$

\section{Numerical results and discussion}

\label{sec4}

We now turn to the numerical evaluation of the sum rules for the coupling
constants. The lower limit of $T$ is determined by the requirement 
that the terms of higher twists in the operator expansion is less 
than one third of the whole sum rule. This leads to $T>1.0$ GeV for
the sum rules (\ref{final-a})-(\ref{final-f}).
In fact the twist-four terms contribute only a few percent to the sum rules 
for such $T$ values. The upper limit of $T$ is constrained by the requirement that
the continuum contribution is less than $30\%$. This corresponds to $T<2.5$GeV.
With the values of pion wave functions 
at $u_0 ={1\over 2}$ given above we obtain the left hand side of  
the sum rules (\ref{final-a})-(\ref{final-f}) as functions of $T$. The
results are shown in FIG. 1.
Stability develops for the sum rule (\ref{final-a})-(\ref{final-f}) 
in the region $1.0$ GeV $<$$T$$<$$2.5$ GeV. Numerically we have:
\begin{eqnarray}
\label{res}
 &&g_1f_{-,{1\over 2} } f_{+, {3\over 2} }
   =-(0.16\pm 0.02\pm 0.03)~~~\mbox{GeV}^{2}\;,\\
 &&g_2f_{-,{1\over 2} } f_{+, {1\over 2} }
   =(0.36\pm 0.02\pm 0.03)~~~\mbox{GeV}^{3}\;,\\
 &&g_3f^2_{+,{1\over 2} }
   =(0.13\pm 0.01\pm 0.02)~~~\mbox{GeV}^{2}\;,\\
 &&g_4f^2_{+,{3\over 2} }
   =(0.10\pm 0.01\pm 0.02)~~~\mbox{GeV}^{2}\;,\\
 &&g_5f_{+,{1\over 2} } f_{+, {3\over 2} }
   =(0.20\pm 0.01\pm 0.02)~~~\mbox{GeV}^{3}\;,\\
 &&g_6f^2_{-,{1\over 2} }
   =(0.20\pm 0.01\pm 0.02)~~~\mbox{GeV}^{2}\;,
\end{eqnarray}
where the first error refers to the variations with $T$  
in this region and the second error takes into account the uncertaity in 
$\omega_c$. And the central value corresponds 
to $T=1.5$GeV and $\omega_c =3.0$GeV for (\ref{final-a})-(\ref{final-e}). 
For the sum rule (\ref{final-f}) we use $\omega_c =2.4$GeV.

With the central values of f's in (\ref{fvalue}) we get the absolute
value of the coupling constants:
\begin{eqnarray}
\label{result}
 &&g_1=-(3.6\pm 0.4\pm 0.6)~~~\mbox{GeV}^{-2}\;,\\
 &&g_2=3.6\pm 0.3\pm 0.6\;,\\
 &&g_3=0.83\pm 0.1\pm 0.2~~~\mbox{GeV}^{-1}\;,\\
 &&g_4=2.7\pm 0.2\pm 0.5~~~\mbox{GeV}^{-3}\;,\\
 &&g_5=2.6\pm 0.2\pm 0.4~~~\mbox{GeV}^{-1}\;,\\
 &&g_6=3.1\pm 0.3\pm 0.6~~~\mbox{GeV}^{-1}\;.
\end{eqnarray}
We have not included the uncertainties due to f's here.
Note that the sum rules (\ref{final-a}) and (\ref{final-c})-(\ref{final-f}) 
are also sensitive to the value of $\varphi_\pi ({1\over 2})$. For example, 
$g_1 =-4.83$ with $\varphi_\pi ({1\over 2})=1.71$ which is one of the values
used in \cite{dyb}.

The couplings of pions with heavy mesons obey symmetry relations 
in the leading order of HQET. Only in HQET the ambiguity due to the presence 
of two distinct $1^+$ states can be resolved. We use light-cone QCD sum rules
to calculate all the six independent strong coupling constants $g_1$-$g_6$ 
for the lowest three doublets. 
The sum rules in HQET for $g_3$ and $g_6$ are consistent with 
those for $g$ and $g^{\prime}$ in \cite{bely95,bari97} in the limit of $m_Q\to \infty$ 
with $g_3={2g^{\prime}\over f_\pi}$ and $g_6={2g\over f_\pi}$.
The sum rule (\ref{final-b}) for $g_2$ can be compared to that 
for $h$ in \cite{bari97} with $g_2={2h\over f_\pi}$ from 
definition. In terms of $h$ our value is $0.25$, which is only one half 
of that given in \cite{bari97}. Our result $g_1=3.6$ is slightly smaller 
than that from the short-distance QCD sum rule \cite{ybd97}.
The calculation for $g_4$ and $g_5$ is new.

With the CLEO measurement of ${\cal B} ({D^\ast}^+ \to D^+ \gamma )$ 
\cite{celo}, a model independent extraction of $g$ has been performed recently 
\cite{iain}. Two possible solutions of $g$ ( $g={f_\pi \over 2} g_6$ in our
notation ) and $\beta$ are found from 
fitting to the experimental data, either $g=0.27^{+0.04+0.03}_{-0.02-0.02}$, 
$\beta =0.85^{+0.2+0.3}_{-0.1-0.1}$GeV$^{-1}$ or $g=0.76^{+0.03+0.2}_{-0.03-0.1}$, 
$\beta =4.90^{+0.3+5.0}_{-0.3-0.7}$GeV$^{-1}$. The $g=0.76$ solution is 
excluded by the experimental limit $\Gamma ({D^\ast}^+ )< 0.13$MeV \cite{acc}.
The coupling $g$ from the LCQSR approach is $g \sim 0.21-0.40$ 
\cite{col94,aliev96,bely95}, which is consistent with the experimentally favored 
solution $g=0.27$. In a recent work \cite{dyb}  
the mixing effect of the two $1^+$ states in the lowest excited heavy 
meson doublets $(2^+, 1^+)$ and $(1^+, 0^+)$ and the ${\cal O}(1/m_Q)$ 
correction to the leading order coupling $g_1$ was calculated. The 
resulting pionic decay widths of $D_2^\ast (2460)$ and $D_1 (2420)$ are 
in agreement with the experimental data \cite{review}.


\vspace{0.8cm} {\it Acknowledgements:\/} S.Z.  was supported by
the National Postdoctoral Science Foundation of China and Y.D. 
was supported by the National Natural Science Foundation of China.

\vspace{1.cm}

{\bf Figure Captions}
\vspace{2ex}
\begin{center}
\begin{minipage}{120mm}
{\sf
FIG. 1.} \small{Dependence of $f_i f_jg_{1-6}$ in (\ref{final-a})-(\ref{final-f})
on the Borel parameter $T$ for 
the continuum threshold $\omega_c =3.0$GeV. The dotted, solid, short-dashed, 
long-dashed, dot-dashed and intermediate-dashed curves correspond to the 
coupling constant $g_1$-$g_6$ respectively.}
\end{minipage}
\end{center}

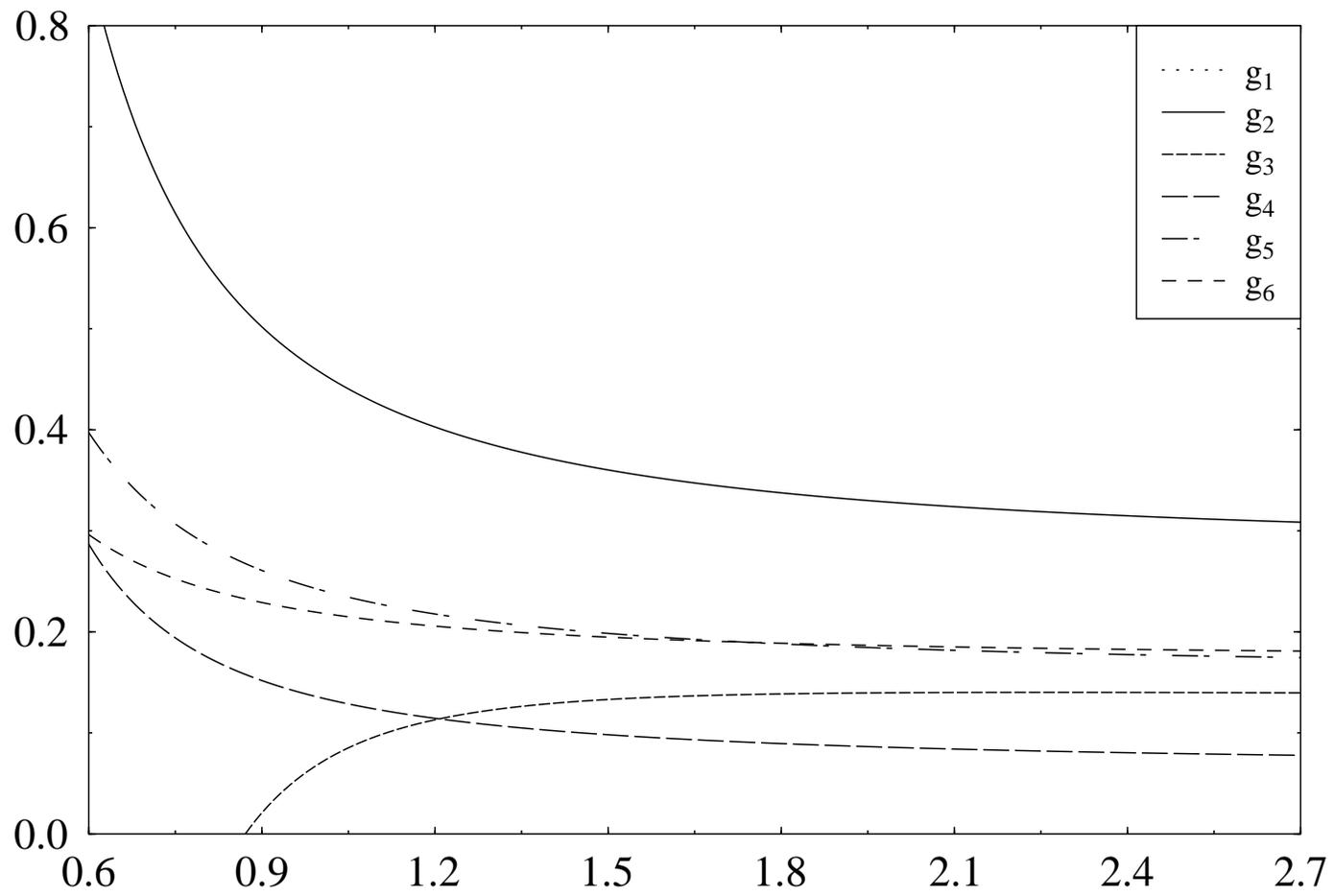
\end{document}